\documentclass{article}
\usepackage{graphics}
\begin{document}
\title{Bubble propagation in a helicoidal molecular chain}
\author{Alessandro Campa\thanks{E-mail: campa@axiss.iss.infn.it} \\
  \small{Physics Laboratory, Istituto Superiore di Sanit\`a} \\
  \small{and INFN Sezione di Roma1, Gruppo Collegato Sanit\`a} \\
  \small{Viale Regina Elena 299, 00161 Roma, Italy}}
\date{July 21, 2000}
\maketitle

\begin{abstract}
We study the propagation of very large amplitude localized excitations in a
model of DNA that takes explicitly into account the helicoidal structure.
These excitations represent the ``transcription bubble'', where the hydrogen
bonds between complementary bases are disrupted, allowing access to the
genetic code. We propose these kind
of excitations in alternative to kinks and breathers.
The model has been introduced by Barbi et al. [Phys. Lett. A
{\bf 253}, 358 (1999)], and up to now it has been used to study on the one
hand low amplitude breather solutions, and on the other hand the DNA melting
transition. We extend the model to include the case of heterogeneous chains,
in order to get closer to a description of real DNA; in fact, the Morse
potential representing the interaction between complementary bases has two
possible depths, one for A-T and one for G-C base pairs. We first compute the
equilibrium configurations of a chain with a degree of uncoiling, and we find
that a static bubble is among them; then we show, by molecular dynamics
simulations, that these bubbles, once generated, can move along the chain.
We find that also in the most unfavourable case, that of a heterogeneous DNA
in the presence of thermal noise, the excitation can travel for well more
$1000$ base pairs.
\end{abstract}

\section{INTRODUCTION}
In the attempt to describe some aspects of DNA functioning, the theory of
nonlinear dynamical systems has found an interesting application to this
important biological structure. In spite of the awareness of the complexity
that characterizes most of the dynamical processes taking place in biological
systems, where very often the necessary trigger is constituted by the
temporary interaction of different objects, there is nevertheless an effort,
in the research in biological physics,
to grasp some essential features with simple models.

In this work we are interested in the propagation of very large amplitude
bubbles, that should be very important in the description of the process of
transcription. In this respect, the
main appeal of onedimensional nonlinear models is their possibility to sustain
localized excitations, of which our bubbles are an example.
In this direction there have been works based
on some models where the
essential degree of freedom of each site of the chain is related to the
opening of the hydrogen bonds between the complementary bases of the double
stranded DNA (for a review see, e.g., \cite{gae,yakb} and references therein).

The important points to address for any given model are the existence and the
stability of localized excitations, both stationary and moving; after that the
problem of their formation has to be considered. In this Section we give a
very brief account of the situation in models with one degree of freedom per
base pair, limiting ourselves to the problems of existence and stability.

Let us begin with homogeneous chains. We can start making
a distinction between models that do not neglect the discreteness of the
system, and models that are treated in the continuum limit, either from the
start or after the approximation of the original discrete equations. In both
cases there is usually a nearest neighbor interaction and a site potential;
in the continuum limit generally the field equation under study is
of the Klein-Gordon type:
\begin{equation}
\frac{\partial^2\phi}{\partial t^2}=v_0^2\frac{\partial^2\phi}{\partial x^2}
-\frac{\partial U}{\partial \phi}
\label{fldeq}
\end{equation}
Consider the problem of the existence of localized stationary solutions.
We know that for equations like (\ref{fldeq})
localized solutions, where at both sides of the excitation the field $\phi$
is at the minimum of $U$, are possible only if $U$ has degenerate minima;
then $\phi$ has a kink configuration and we have a topological soliton,
which implies a displacement of a whole side of the chain. One of the most
used example of (\ref{fldeq}) is the sine-Gordon equation. The stationary
solutions really are static, i.e., they are equilibrium configurations, that
are obtained by solving the Newton-like
equation $v_0^2\frac{\partial^2\phi}{\partial x^2}=\frac{\partial U}
{\partial \phi}$, and localized solutions, the kinks, are found by choosing
appropriate ``initial conditions'' in this equation. If the original model
equations were discrete, the equilibrium configurations are such that their
envelope has a kink structure which is very close to that of the continuum
equation, with the center exactly on one site (odd kink), or in the middle
between two sites (even kinks) (see \cite{bope}, where also the problem of
the stability is considered).

In models where the discreteness is taken into account, the problem of
solutions with a topological index can be circumvented; in fact, also with site
potentials $U$ with a nondegenerate absolute minimum, it is possible to have
stationary localized excitations, in the form of breathers, in which only few
sites are coherently involved with a not negligible amplitude in a nonlinear
oscillation with a given fundamental frequency (for the mathematics of
breathers in discrete nonlinear lattices and the conditions for their
existence and stability see \cite{mac,aub,fla}). Now there is an additional
degree of freedom in the choice of the localized solution, namely the
variation, in proper ranges, of the fundamental frequency.
In analogy with the
situation that arises with kinks, a breather with a given fundamental
frequency can be centered on one site, which has the largest amplitude of
oscillation (odd breather), or on two sites, which have the largest identical
amplitudes (even breather).

We now turn our attention to the stability of these solutions.
The stability of a static kink configuration, in lattice models approximated
by (\ref{fldeq}), requires the
positive definiteness of the hessian matrix of the total potential energy.
Instead for a breather the stability is checked through the linearization
of the equations of motion around the stationary solution; the transformation
of the perturbation in one breather period gives a linear application,
whose spectrum determines the stability properties:
the breather is stable if there are no eigenvalues with modulus greater than
$1$ \cite{aub}.

The important point to study is that of the moving capabilities of localized
excitations.
If we are interested in moving solutions of Eq. (\ref{fldeq}), the problem
is easily solved. Any static solution, in particular a localized one,
can be transformed to a moving one, with speed $v<v_0$, through a Lorentz
boost (with ``speed of light'' $v_0$); its profile will only be modified
by the Lorentz contraction. If the original model is discrete and
Eq. (\ref{fldeq}) is only its approximation, then the movement of the localized
excitation will be associated to an energy loss through phonon radiation
but generally the kink retains a good stability. In
\cite{boes} one can find the treatment of this phonon dressing of the moving
kink.

The study of moving breathers is more difficult. While the existence of
stationary breathers seems to be quite independent of the characteristics
of the site potential, at least for sufficiently small coupling between
different sites \cite{aub}, the presence of exact moving breathers is
associated to some special integrable hamiltonians (see \cite{fla} for a
review). In the other, generic cases, the common approach is to
study the stability spectrum of stationary breathers, making a connection
between some of the elements of the spectrum and the ``sensitivity'' to
movement \cite{aucr,fla1}. The absence of a
zeroth order moving solution, analogous to the boosted kink of Eq.
(\ref{fldeq}), makes the treatment of stability of moving breathers less
approachable.

If we consider heterogeneous chains, there are of course more problems,
especially regarding the stability of moving excitations, in particular
breathers. The amount of work has been less extensive. There have been studies
on nonlinear localization in chains with impurities \cite{kivs}, on kink
propagation through regions with mass variation \cite{kalo}, MD simulations
in a sine-Gordon model of DNA \cite{sal}; see \cite{gre} for a review of
results on disordered systems in the continuum limit.

Let us summarize what are the main problems with kinks and breathers if we
want to consider them as good candidates for the local openings needed during
the transcriptions. As we have seen, the stability is a less severe difficulty
for kinks. For this reason they could be preferable. On the contrary, the
breathers have the advantage to avoid the problem of the topological index.
However, it is plausible that, if we want to describe with a breather a
region where the bond between the complementary bases is temporarily
broken, to allow access to the genetic code, this breather must be wide
(i.e., the number of sites essentially involved in the motion is not very
small) and of very large amplitude; unfortunately, the probability that a
breather is stable greatly decreases when its width increases and when its
amplitude is large \cite{aub} (this
second point can be easily guessed, since for potentials that allow
dissociation, large amplitude means low fundamental frequency, and therefore
a strong probability of resonance with the phonon dispersion curve).

In this work it is our purpose to show that, with a given model with two
degrees of freedom per base pair, it is possible to put together some of the
advantages of kinks and breathers, i.e., respectively, a good stability with
respect to movement, and a local excitation that does not need a
topological index. We will show that these large amplitude solutions of the
model have a satisfying stability also with heterogeneity and with thermal
noise. We think that these properties can represent those of a
``transcription bubble''.

The model has been proposed by Barbi et al. \cite{barb},
and it is an evolution, that takes the helicoidal structure
explicitly into account, of the Peyrard-Bishop model. This last model was
introduced \cite{pb89} to have a dynamical explanation of the melting
transition, opposed to methods that offer only equilibrium estimates of the
probability of bond disruption \cite{pol,wart,pro}. A satisfactory melting
curve was obtained \cite{pb93}, and later the melting of heterogeneous
chains and of heterogeneous oligonucleotides has been studied
\cite{cul,zhan,ca1,ca2}. The helicoidal model introduced in \cite{barb} was
there used to build approximate low amplitude solutions through the
method of the multiscale expansion \cite{mult}, and in \cite{cocc} the
melting transition was investigated.

In Section II we introduce the model and we compute its equilibrium
configurations, with their stability properties. In Section III we show the
results of our MD simulations, together with an approximate computation of the
features of the moving bubbles. In Section IV we present our discussion and
draw some conclusions.

\section{THE MODEL}\label{model}
Our starting point is the model introduced (in a somewhat different version)
in \cite{barb}.
The bases can move only in planes perpendicular to the helix axis; besides,
the center of mass of the base pair is held fixed, and the two complementary
bases move simmetrically with respect to the axis of the molecule. Then for
each base pair there are two degrees of freedom:
$r_n$ is the distance between each one of the complementary bases in the
$n$-th base pair and the helix axis; $\theta_n$ is the angle that the line
joining the two complementary bases makes with a given direction in the
planes where the bases move. The potential energy is given by:
\begin{eqnarray}
U&=\sum_n \Bigl\{& D_n{\left(e^{-a(r_n-R_0)}-1\right)}^2+\frac{1}{2}c
{(r_{n+1}-r_n)}^2 \nonumber \\
&&+\frac{1}{2}K{\left[ L_{n+1,n}-L_0\right]}^2\Bigr\}
\label{enpot}
\end{eqnarray}
where $L_{n+1,n}$ is the distance between neighbor bases on the same strand,
and as a function
of $r_n$, $r_{n+1}$ and $\theta_{n+1}-\theta_n \equiv \Delta\theta_n$ is given
by:
\begin{equation}
L_{n+1,n}= \sqrt{h^2+r_{n+1}^2+r_n^2 -2r_{n+1}r_n\cos^2\Delta\theta_n}
\label{distfun}
\end{equation}
where $h=3.4$ \AA \, is the fixed distance between neighbor base planes and
$R_0=10$ \AA \, is the equilibrium value of $r_n$; $L_0$
is the same function computed for $r_{n+1}=r_n=R_0$ and $\Delta\theta_n
=\Theta_0=\pi/5$ ($10$ base pairs per helix turn). Therefore
the equilibrium configuration is that with
$r_n=R_0$ and $\Delta\theta_n=\Theta_0$ for each $n$,
which gives the system its helicoidal
structure. The natural helicity is right handed, and for convenience we take
this as the positive rotation for the angles $\theta_n$. Actually,
equilibrium configurations are also all those in which any $\Delta\theta_n$
is chosen indifferently $\pm \Theta_0$; however, they are separated
by the fundamental one by potential barriers, that in our simulations are
never crossed.
The first two terms
in (\ref{enpot}) are the same of the original
PB model \cite{pb89}, which has $r_n$ as the only degree of freedom per base
pair: there is a
Morse potential with $R_0$ as the equilibrium distance, and a harmonic
interaction between neighbor base pairs (stacking interaction);
there can be two different values
for $D_n$, $D_{A-T}$ for A-T base pairs and $D_{G-C}$ for G-C base pairs. It
is generally assumed that $D_{G-C}=\frac{3}{2}D_{A-T}$. The
last term in (\ref{enpot})
describes a restoring force that acts when the distance $L$ between neighbor
bases on the same strand is different from $L_0$.
The model has been introduced in \cite{barb} without the second term in
(\ref{enpot}), essentially attributing all the stacking interaction to the last
term, and with an additional three-body term proportional to
${(\theta_{n+1}+\theta_{n-1}-2\theta_n)}^2$, to eliminate the equilibrium
configurations with some $\Delta\theta_n=-\Theta_0$. In this
form the authors
have studied small amplitude breather-like solutions, with the envelope
described by the nonlinear Schr\"{o}dinger equation. In \cite{cocc} the
statistical mechanics of model (\ref{enpot}) has been studied, the authors
being interested in
the melting transition of DNA; in this case the difference from (\ref{enpot})
was given by a replacement of the coupling constant $c$ by a coupling of the
form $ce^{-b(r_{n+1}+r_n-2R_0)}$, decreasing with $r_{n+1}+r_n$ increasing.
Besides, the restoring force represented by the last term was cast in another
form, with $L$ fixed ($\equiv L_0$) and $h$ variable.
In both works a homogeneous DNA ($D_n$ constant in $n$) was considered. We use
the more complete structure used in \cite{cocc}, with the second term in
(\ref{enpot}) more related to the stacking interaction, and the last term
more related to the rigidity of the two single strands,
with $L$ variable and $h$ fixed. For simplicity, we do not insert a decaying
coupling: this last feature has been found,
already in the original PB model \cite{pb93}, as being essentially responsible
for
sharpening the melting transition, which happens at higher temperatures than
those we are interested in here. Also, the three-body term used in
\cite{barb} was not necessary, since in our simulations, as we have already
pointed out, we never had a
crossover in the sign of $\Delta\theta_n$.

The spirit in which we study the model represented by (\ref{enpot}) is the
following. We consider a chain of a given length, with fixed boundary
conditions. We know that an interaction with an enzymatic complex (with RNA
polymerase) is necessary to trigger the process of transcription.
We take here, as a working hypothesis, that this external action can be
represented by a partial unwinding of one of the extremes of our chain,
considered as the interaction site. We will show that this mechanism can give
rise to a travelling bubble (the ``transcription bubble''), in which several
base pairs are open; this bubble, that appears to be very stable in a
homogeneous chain initially at rest, is interestingly long lived also in the
case of a heterogeneous chain and with thermal disorder.

In the remaining of this Section our aim is to give an analytical
background to the results, obtained by molecular dynamics simulations, that
will be presented in the next Section. Since the essential dynamical process
that we will observe is a local opening travelling along the chain, we want to
show that this movement can be considered, in an adiabatic approximation, as
a succession of equilibrium configurations, similarly to what happens with
the travelling of kinks. Consequently, in the first subsection we will
study the equilibrium configurations of our system: we will show how a chain
with some uncoiling, caused by suitable boundary conditions
(i.e., if $\theta_0$ and $\theta_{N+1}$ are held fixed at values such that
$\theta_{N+1}-\theta_0 \equiv \sum_{n=0}^N\Delta\theta_n < (N+1)\Theta_0$),
can have different equilibrium configurations. The simplest one, for a
homogeneous chain, is given by a homogeneous configuration $r_n=r$ and
$\Delta\theta_n=\Delta\theta$ for each $n$, for certain values of $r$ and
$\Delta\theta$; for a heterogeneous chain, where $D_n=D_{A-T}$ for some $n$
and $D_n=D_{G-C}$ for the other values of $n$, the configuration is not
qualitatively very
different from the previous, although the precise equilibrium values of
$r_n$ and $\Delta\theta_n$ depend on the sequence. Another
equilibrium configuration in the homogeneous case, the one in which we are
interested, is one in which a small
region (about $20$ base pairs) is completely open, and at both sides $r_n$
and $\Delta\theta_n$ decay rapidly to homogeneous values.
Again, in the heterogeneous case, the dependence on the sequence
does not alter qualitatively the picture. In the second subsection we will
briefly treat the stability of these equilibrium configurations.

\subsection{Equilibrium configurations}\label{subeqcn}
If we neglect the mass variation between A-T and G-C base pairs, and take a
proper unit of mass, the equations of motion deriving from (\ref{enpot}) are:
\begin{eqnarray}
\ddot{r}_n-r_n\dot{\theta}_n^2&=&-\frac{\partial U}{\partial r_n} \nonumber \\
r_n^2\ddot{\theta}_n+2r_n\dot{r}_n\dot{\theta}_n&=&-\frac{\partial U}
{\partial \theta_n}
\label{eqmot}
\end{eqnarray}
The equilibrium configurations are those that make the right hand sides vanish.
Therefore we have to solve:
\begin{eqnarray}
2aD_n\left(e^{-2a(r_n-R_0)}-e^{-a(r_n-R_0)}\right)+c\Delta_2r_n&&
\nonumber \\
+K\Biggl\{\frac{L_{n+1,n}-L_0}{L_{n+1,n}}\left[r_{n+1}\cos\Delta\theta_n
-r_n\right] \phantom{xxxxxx}&& \nonumber \\
+\frac{L_{n,n-1}-L_0}{L_{n,n-1}}\left[r_{n-1}\cos\Delta\theta_{n-1}
-r_n\right]\Biggr\}=0&&
\label{eqconfa}
\end{eqnarray}
\begin{eqnarray}
\frac{L_{n+1,n}-L_0}{L_{n+1,n}}r_{n+1}r_n\sin\Delta\theta_n
\phantom{xxxxxx}&&\nonumber \\
-\frac{L_{n,n-1}-L_0}{L_{n,n-1}}r_{n-1}r_n\sin\Delta\theta_{n-1}=0&&
\label{eqconfb}
\end{eqnarray}
for $n=1,\ldots,N$; here $\Delta_2r_n\equiv r_{n+1}+r_{n-1}-2r_n$. From the
structure of
Eq.\ (\ref{eqconfb}) it is clear that any solution of (\ref{eqconfa}) and
(\ref{eqconfb}) has to
be such that the quantity represented by, say, the first term in
(\ref{eqconfb}) is a constant as a function of $n$. Let us begin considering
a homogeneous chain. Then of course the simplest
solution is to have both $r_n$ and $\Delta\theta_n$ constant in $n$. In this
case also $L_{n+1,n}$ is constant in $n$. Therefore, posing $r_n=r$ and
$\Delta\theta_n=\Delta\theta$, and equating $L_{n+1,n}$ to a constant
$\overline{L}$, one can express $\Delta\theta$ as a function of $r$ and
$\overline{L}$; then, substituting in (\ref{eqconfa}) this function, together
with $L_{n+1,n}=\overline{L}$ and $r_n=r$, one can find (numerically) the
equilibrium value $r$ (which will depend on $\overline{L}$). Going back also
the equilibrium value $\Delta\theta$ can be computed. If the chain is not
infinite and there are fixed boundary conditions (for $n=0$ and $n=N+1$),
then, for the equilibrium it is required that also $r_0=r_{N+1}=r$ and
$\Delta\theta_0=\Delta\theta_N=\Delta\theta$. It is clear that the fundamental
equilibrium configuration ($r=R_0$ and $\Delta\theta= \Theta_0$) is obtained
for $\overline{L}=L_0$. For $\overline{L}<L_0$ we have $r>R_0$ and
$\Delta\theta<\Theta_0$ (uncoiling), while for $\overline{L}>L_0$ we have
$r<R_0$ and $\Delta\theta>\Theta_0$ (overcoiling). When the chain is
heterogeneous, the corresponding equilibrium solution can be found from the
homogeneous one with an iterative
procedure explained in Appendix A. The solution will depend on the sequence
of the $D_n$s; however, it will not be qualitatively very different from the
homogeneous case. The interesting equilibrium configuration is of course
that in which we have an open region. In this case, although
the first term in (\ref{eqconfb}) is constant in $n$ (and equal, say, to $s$),
$L_{n+1,n}$ is not itself a constant.
We have developed a procedure to compute these configurations. Here we only
give a sketch; more details can be found in Appendix B. In principle, from
\begin{equation}
(L_{n+1,n}-L_0)r_{n+1}r_n\sin\Delta\theta_n=sL_{n+1,n}
\label{forbub}
\end{equation}
it is possible
to obtain $\Delta\theta_n$ as a function of $r_{n+1}$, $r_n$ and $s$;
substituting in (\ref{eqconfa}), we can therefore obtain $r_{n+1}$ as a
function of $r_n$, $r_{n-1}$ and $s$; in this way, starting from
the values for two contiguous $r_n$, we can compute site by site the
equilibrium configuration. With a proper choice of the value of $s$, we
obtain a solution in which there is a region, of about $20$ base pairs, where
$r_n>R_0$ in such a way to stay in the plateau of the Morse potential; in
that region the uncoiling ($\Delta\theta_n<\Theta_0$) is marked. At both
sides of the open region the $r_n$s decay very rapidly to a value slightly
larger than $R_0$ and the $\Delta\theta_n$s to a value slightly smaller
than $\Theta_0$. As before, after the computation has been performed for
a homogeneous chain, with the procedure
explained in Appendix A we can obtain also the configuration for a
heterogeneous chain. The two cases do not differ qualitatively. In Fig. 1
we show two examples of equilibrium configurations: one for a
homogeneous chain and another for a heterogeneous chain, in which the
sequence of A-T and G-C has been chosen randomly; we present only the graphs
concerning the radial degrees of freedom $r_n$.

In the homogeneous case,
the configuration is symmetric at both sides of the bubble. Besides the obvious
translational invariance (for the infinite chain), it is possible to have
a configuration centered on one site with the largest opening, as in
Fig. 1, or on two sites with equal and largest openings (the
analogous of what happens also for the discrete kinks and breathers). For
brevity, in the following the bubble centered on one site will be denoted
odd bubble, and that centered on two sites even bubble. In the
heterogeneous case, translational invariance is lost, but it is not difficult
to guess, in view of the method described in Appendix A, that an equilibrium
configuration occurs for any site or couple of neighbor sites chosen as the
center of the bubble (obviously now not symmetric).

\subsection{Stability}\label{substcn}
In order for an equilibrium configuration to be stable, the hessian matrix
of the potential $U$ must be positive definite at that point of
configuration space. In that case, the square roots of twice the eigenvalues
of the matrix give the proper frequencies of the small oscillations around
the equilibrium. We will consider here, as an example, the results
for $s=-0.273$ (see Eq. (\ref{forbub})),
for the cases of the odd and the even bubble. However, before treating
our particular example,
we want to note the following fact concerning the stability of these kind
of configurations.
We have found that, depending on the choice of the constant $s$ and on the
values of the model parameters, both stable and unstable cases occur, and often
if the odd bubble is stable, then the even bubble
is unstable, and viceversa. Then
one of the smallest eigenvalues of the hessian matrix in the
stable case is associated to the movement of the bubble
along the chain; to be more precise, the perturbation that
corresponds to the eigenvector associated to this small eigenvalue gives
rise, once the linear regime has been passed, to the movement of the bubble
in one direction. During the movement the bubble will go (in the adiabatic
approximation) through the equilibrium configurations constituted by the
odd and the even bubbles. We have here a strong similarity with the situation
that arises with kinks \cite{boes}, and an analogy with the breather
sensitivity to movement that we mentioned in the Introduction.

We now turn to our example with $s=-0.273$ and with the same parameter values
that we employ for the simulations (we will give these values at the beginning
of the next Section).
We have performed our calculations on a chain
with $100$ base pairs, with the bubble in the middle; this should be
sufficient to avoid boundary effects.
For the odd
bubble the hessian matrix is positive definite. Most of the proper frequencies
are associated to phononic excitations; in fact, the corresponding eigenvectors
are spread throughout the all chain. But a small number of eigenvalues
correspond to eigenvectors that have components which are not negligible
only on the sites of the open region. Therefore, they represent perturbations
of the bubble. Among these, there is the smallest eigenfrequency \cite{not1}.
In Fig. 2 we show the eigenvector corresponding to the smallest eigenvalue.
This is clearly associated to the movement
of the bubble, according to what we pointed out in the previous paragraph.
This is proved by the fact that in the spectrum of the even bubble the
eigenvalues are all positive except one. The positive eigenvalues are very
similar to the corresponding ones of the odd bubbles, while the negative one
is very close, in absolute value, to the smallest of the odd bubble. The
smallness of this value shows that the open region is very ``sensitive''
to movement \cite{aucr}.

\section{RESULTS}
In this Section we show the results of the simulations performed. We have
simulated a chain of $2500$ base pairs, with fixed boundary conditions.
The parameters of the model have been given the following values:
the depths of the Morse potential
are $D_{A-T}=0.05$ eV and $D_{G-C}=0.075$ eV; the width is $a=4$ \AA$^{-1}$;
the constant of the harmonic stacking interaction is $c=0.05$ eV/\AA$^2$,
while that of the restoring force is $K=0.14$ eV/\AA$^2$; we have already
given the values of $h=3.4$ \AA, $R_0=10$ \AA \, and $\Theta_0=\pi/5$.
We have used the second order bilateral simplectic algorithm described in
\cite{case}.
As anticipated before, the travelling bubble is formed by imposing a partial
unwinding at one end of the chain. After that, the open region travels
towards the other end. Let us give some more details on the process by
which the open region is formed. We have fixed boundary conditions. At the
left end of the chain we begin to make an
unwinding. This is done by decreasing the angle $\Delta\theta_0=\theta_1
-\theta_0$ between
the ``fixed'' site at the left of the chain and the first site, i.e., by
increasing $\theta_0$. This causes
an opening of the first few sites because of the last term in the potential
energy (\ref{enpot}). During the process of formation of the open region,
also phonons are created, which begin to travel faster than the bubble. At
the end of the process we observe the bubble travelling towards the right.
We have used different amplitudes for the increase of $\theta_0$, that will
be specified in the following for the different cases.

We begin by showing the results of the simulation performed for a homogeneous
chain initially at rest in the fundamental equilibrium configuration. We have
increased $\theta_0$ by $1.25$ radians. In Fig. 3 we present the configurations
at $6$ different times. It can be noted from
the graphs that the travelling bubble is practically stable. We have even found
that, when it reaches the end of the chain, it bounces back. Another thing to
be noted, and that is valid also in the other situations that we will
show, is that outside the bubble the radial coordinate $r_n$ is practically
in the equilibrium position $R_0$, and correspondingly there is no
uncoiling. This appears to be in
contrast with the results of the previous Section, concerning the structure
of the equilibrium configurations. We remind that we found a degree of
uncoiling, and a value of $r_n$ somewhat larger than $R_0$, at the sides of
the bubble. At the end of this Section we will try to give an argument to
show the reason why the sides of a travelling bubble can be in the
equilibrium configuration. This fact is probably a good point,
in view of a possible biological significance of
the dynamical processes of this model, and we will comment on that in the last
Section.

In Fig. 4 we show the situation that arises in a heterogeneous chain, again
initially at rest in the fundamental state. The sequence of base pairs has
been chosen at random. We can note that the bubble progressively decreases
its amplitude, contrary to what happens in the homogeneous case, and in fact in
the last configuration practically we do not see it any more. However,
before disappearing the excitation has travelled well beyond $1000$ base
pairs. Here we have increased $\theta_0$ by a greater quantity than
before, namely by $2$ radians.
It is not difficult to understand the reason of the different
behavior between homogeneous and heterogeneous chains. In the first case the
spectrum of the hessian matrix in the equilibrium positions for given $s$
(see Eq. (\ref{forbub})) is the same for all odd bubbles and the same for all
even bubbles, indipendently of the location (at least for infinite chains, but
for finite chains this is true to a high degree of accuracy, unless the bubble
is very close to one end of the chain). Therefore, in the adiabatic
approximation, the dynamical situation of a bubble repeats periodically
every site that has been travelled. In a heterogeneous chain the hessian
matrix is, in general, different at any location, thus the above argument
does not apply, and a faster energy loss takes
place. Nevertheless the lifetime of the bubble is still satisfying. It is
possible to argue, in a qualitative way, that heterogeneity acts on the
bubble only through the few sites belonging to its two ends, since the
other sites of the bubble are in the plateau region of the Morse potential,
where there are no differences between the two types of base pairs. With the
exposition of the results obtained for chains at room temperatures, we will
touch again this point.

We have made simulations in which we have produced, with the same
procedure as before, a localized excitation; but now the chain is initially
in thermal equilibrium at $300$ K. Again, we have studied both a
homogeneous and a heterogeneous chain. For the second case, we have used the
same base pair sequence that has been adopted for the simulation of the
system initially at rest. Let us first consider the homogeneous chain. We have
made a simulation in which we have increased $\theta_0$ by the same quantity,
$2$ radians, used in the heterogeneous chain at zero temperature. In this
way we could make a comparison, under the same initial excitation process,
between the robustness of the bubble against heterogeneity and against
this level of thermal noise. Fig. 5 shows the configurations again at $6$
successive times. From the inspection of Figures 4 and 5 we can note the
following points. The amplitude of the bubble is greater in the heterogeneous
chain initially at rest; nevertheless the distance travelled is somewhat
greater in the homogeneous chain at $300$ K. Therefore it seems that the
interaction with the phonon bath at this temperature is less effective, in
taking energy away from the bubble, than the modes of the (disordered)
heterogeneous chain. Of course, it is possible to increase the lifetime of
a bubble by increasing the strength of the initial excitation. In Fig. 6
the configurations obtained for the omogeneous thermalized chain when
$\theta_0$ is initially increased by $2.8$ radians. We can see that the bubble
has still a large amplitude when it has almost reached the end of the chain;
as in the case of the zero temperature, we have found that it bounces back.

The last case that we present is that of the heterogeneous chain at $300$ K,
in Fig. 7; the initial increase in $\theta_0$ is $2.8$ radians. We see that,
in spite of the two possible sources of disturbances to the localized
excitation, heterogeneity and thermal noise, the bubble still travels for
about $1300$ base pairs.

\subsection{Moving open regions}\label{submvcn}
In order to show how the bubbles move along the chain, we employ here a
simplified version of the model. We
make this choice since in this way we can have manageable expressions. However,
we believe that the same kind of mechanisms happen in the complete system,
the difference being that the expressions would be much more involved,
requiring the inversion of trigonometric functions.

The fundamental equilibrium configuration is that with $r_n=R_0$ and
$\theta_n=n\Theta_0$; we here expand $L_{n+1,n}$ (see Eq. (\ref{distfun}))
in power series and keep
only the first order terms in $(r_n-R_0)$, $(r_{n+1}-R_0)$ and
$(\theta_n -n\Theta_0)$. Such a procedure is not entirely
consistent, since we do not make an analogous expansion in the Morse potential.
However, we have checked numerically that the linear approximation for
$L_{n+1,n}$ is not bad in a quite large range of variability of $r_n$,
$r_{n+1}$ and $\Delta\theta_n$, and more importantly this simplification
is done only for illustrative purposes. Calling
$y_n\equiv r_n-R_0$ and $z_n\equiv R_0(\theta_n-n\Theta_0)$, and neglecting
the kinetic terms, the equations of motion of this simplified system
in the homogeneous case are:
\begin{eqnarray}
\ddot{y}_n&=&2aD\left(e^{-2ay_n}-e^{-ay_n}\right)+c(y_{n+1}+y_{n-1}-2y_n)
\nonumber \\ &-&A^2(y_{n+1}+y_{n-1}+2y_n)-AB(z_{n+1}-z_{n-1})
\label{eqsimpa}
\end{eqnarray}
\begin{equation}
\ddot{z}_n=B^2(z_{n+1}+z_{n-1}-2z_n)+AB(y_{n+1}-y_{n-1})
\label{eqsimpb}
\end{equation}
where the positive coefficients A and B, coming from the power expansion of
$L_{n+1,n}$, are given by $A=2\sqrt{K}\frac{R_0}{L_0}\sin^2\frac{\Theta_0}{2}$
and $B=\sqrt{K}\frac{R_0}{L_0}\sin\Theta_0$. Going to the continuum limit,
we pose
$n\rightarrow x$, $y\rightarrow \phi$ and $z\rightarrow \psi$. Taking into
account partial spatial derivatives up to a suitable degree, and denoting
with $U_M$ the Morse potential, we obtain the following equations:
\begin{eqnarray}
\frac{\partial^2\phi}{\partial t^2}&=&-\frac{\partial U_M}{\partial \phi}
-4A^2\phi -A^2\frac{\partial^2\phi}{\partial x^2} \nonumber \\ &+&
c\frac{\partial^2\phi}{\partial x^2} -2AB\frac{\partial \psi}{\partial x}
-\frac{1}{3}AB\frac{\partial^3 \psi}{\partial x^3}
\label{eqconta}
\end{eqnarray}
\begin{equation}
\frac{\partial^2\psi}{\partial t^2}= B^2\frac{\partial^2\psi}{\partial x^2}
+\frac{1}{12}B^2\frac{\partial^4\psi}{\partial x^4} +2AB\frac{\partial \phi}
{\partial x} +\frac{1}{3}AB\frac{\partial^3 \phi}{\partial x^3}
\label{eqcontb}
\end{equation}
We now pose
$\frac{\partial^2\phi}{\partial t^2}=v^2\frac{\partial^2\phi}{\partial x^2}$
and
$\frac{\partial^2\psi}{\partial t^2}=v^2\frac{\partial^2\psi}{\partial x^2}$,
and in the following we consider the expressions that are found keeping
only the first order terms in $v^2$.
From Eq. (\ref{eqcontb}) it is possible to obtain an expression for
the spatial derivatives of $\psi$ as a function of $\phi$; the one that is
of interest to us is:
\begin{equation}
\frac{\partial \psi}{\partial x}= d\left( 1+\frac{v^2}{B^2}\right)
-2\frac{A}{B}\left(1+\frac{v^2}{B^2}\right)\phi
-\frac{A}{6B}\frac{\partial^2 \phi}{\partial x^2}
\label{eqpsi}
\end{equation}
with the arbitrary constant $d$.
Substituting in (\ref{eqconta}) for $\frac{\partial \psi}{\partial x}$ and
$\frac{\partial^3 \psi}{\partial x^3}$ we find an equation for $\phi$:
\begin{equation}
c'\frac{\partial^2\phi}{\partial x^2}=-\frac{\partial}{\partial \phi}
\left[-U_M(\phi)+g\left(1+\frac{v^2}{B^2}\right)\phi +2\frac{A^2}{B^2}
v^2\phi^2\right]
\label{eqtrav}
\end{equation}
where $c'=c-v^2\left(1-\frac{2A^2}{3B^2}\right)$ and $g=-2dAB$. Let us begin
considering the static case, $v^2=0$. Then, Eq. (\ref{eqtrav}) reduces to:
\begin{equation}
c\frac{\partial^2\phi}{\partial x^2}=-\frac{\partial}{\partial \phi}
\left[-U_M(\phi)+g\phi\right]\equiv -\frac{\partial}{\partial \phi}V(\phi)
\label{confcont}
\end{equation}
We see that with a positive $g$ (i.e., with $d<0$) we have the
possibility of a localized excitation; actually, it is not difficult to
see that it must be $0<g<\frac{1}{2}aD$. In fact, in that case $V(\phi)$,
that diverges exponentially to $-\infty$ for $\phi\rightarrow-\infty$ and
linearly to $+\infty$ for $\phi\rightarrow+\infty$,
has a local maximum for a (small) positive value $\phi^*$ and a local
minimum for a larger value of $\phi$. These two values are given by the two
solutions of the equation $\frac{\partial}{\partial \phi}U_M=g$. Then, solving
the Newton-like equation (\ref{confcont}) with a ``total energy''
$V(\phi^*)$, we have either $\phi\equiv\phi^*$ or $\phi\rightarrow\phi^*$
for $x\rightarrow\pm\infty$, with a localized region where $\phi$ reaches
a maximum; this maximum is given by the value of $\phi$, to the right of the
local minimum, where $V(\phi)=V(\phi^*)$. In this case, at the sides of the
open region we also have
$\frac{\partial \psi}{\partial x}\rightarrow d-2\frac{A}{B}\phi^*<0$, i.e.,
a small uncoiling. Summarizing, in order to have a static localized
solution a constant $d<0$ is necessary.

We now go to $v^2>0$ (although sufficiently small since we have made an
expansion in $v^2$). We rewrite Eq. (\ref{eqtrav}) with $g=0$ (i.e., $d=0$):
\begin{equation}
c'\frac{\partial^2\phi}{\partial x^2}=-\frac{\partial}{\partial \phi}
\left[-U_M(\phi)+2\frac{A^2}{B^2}v^2\phi^2\right]
\equiv -\frac{\partial}{\partial \phi}W(\phi)
\label{eqtravg}
\end{equation}
The differences from before are that the divergence of $V(\phi)$ for
$\phi\rightarrow+\infty$ is now quadratic, and, more important for our
argument, the local maximum is for $\phi=0$. Therefore, it is
possible to have travelling localized excitations, at both sides of which the
field $\phi$ goes to $0$, and so does the uncoiling $\frac{\partial \psi}
{\partial x}$.

Although we have used here a simplified model, it is very likely that in the
complete system a very similar argument applies. This should explain why in
the simulations we find, at the sides of the open region, normal twisting.

\section{DISCUSSION AND CONCLUSIONS}
In this work we have studied a model of DNA with two degrees of freedom per
base pair. The model has been built explicitly to represent the helicoidal
structure of DNA \cite{barb}. We have analytically shown that, under some
uncoiling, the system exhibits stable equilibrium configurations in which
there is a small region, of about $20$ base pairs, where the hydrogen bond
between complementary bases is completely disrupted, allowing access to the
genetic code. Then, through MD simulations, we have found that these
open regions can travel along the DNA chain, also when both thermal noise and
heterogeneity are present.

In connection with our results,
we would like to mention what has been found in \cite{cocc} concerning the
statistical mechanics of this model (the small differences in the Hamiltonian
of \cite{cocc} should not be important for this qualitative point). In that
work the melting transition has been studied. The isothermals in the plane
with the thermodynamical variables corresponding to torque and uncoiling show
clearly a first order phase transition (the computation are performed for an
infinite chain); during the transition, in which the uncoiling increases at
constant temperature and torque, the two coexisting phases are
interpreted as one with normal distance between the complementary bases, and
one with the hydrogen bonds disrupted. At the end of the transition, only the
phase with disrupted bonds remains. It is natural to think that these two
phases can be put in correspondence with the two possible equilibrium
configurations that exist in a chain with a degree of uncoiling, namely that
with a bubble and that without, taking into account that our simulations are
performed at a temperature (or at a torque) below that required for the
melting transition.

We have noted in the previous Section that the travelling bubbles that have
been generated in our simulations show normal coiling at the sides of the
open region, and in correspondence the hydrogen bond between complementary
bases is at the equilibrium distance. This suggests the possibility to have
more than one travelling bubble at the same time. This fact resembles the
situation that arises with kinks: only one static kink can be present (and
this is easy to understand, since the exact solution reaches the positions of
the minimum of the potential only asymptotically), but for travelling kinks
the situation is different \cite{raja}. Therefore, this model allows
transcription to take place at the same moment in different portions of the
chain.

In the construction of nonlinear dynamical models of biological systems, one
of the main properties to satisfy is robustness of the relevant processes.
This means that changes, at least within suitable ranges, of the external
conditions, or of the dynamics of the triggering events, or even of the
parameters of the effective potentials, must not result in essential changes
of the main features of the process under consideration. The topological
index of kinks can not be destroyed by perturbations or by thermal
disorder, although, of course, a kink can loose energy by phonon radiation
\cite{boes}. Breathers are non topological objects,
but some simulations \cite{dau1} have shown that they can survive
perturbations. However, the larger their energy, the larger their tendency to
remain pinned \cite{dau1,dau2};
besides, as we mentioned in the Introduction, if we look for a breather with a
very large amplitude, as should be required to allow exposition of the genetic
code, then we will not find a stable excitation.

The model used in this work, with two degrees of freedom per base pair, has
shown to possess the positive features of both kinks and breathers: although
there is no topological index to prevent eventual decay of the excitation,
nevertheless the ``transcription bubble'' is quite stable. A necessary
condition for biological plausibility is that heterogeneity must not prevent
propagation of the localized excitations in this class of models. We have
seen that this is our case, also if certainly the life of the bubbles is
shorter if the chain is not homogeneous. We have argued that this is due to
the very structure of the bubble: most of the sites belonging to it are in
the plateau region of the Morse potential, where there are no differences
between base pairs; only the few sites at the two ends of the bubble
experience these differences.

We have chosen to
generate the open region through an unwinding at one of the ends of the chain;
this should simulate the initial enzymatic action. We would like to say more
on the spirit in which this position has been taken. There have been attempts
to see how breathers can form spontaneously during the dynamics,
starting from modulational instability, and then growing through
collisions, that on the average favor the growth of the larger excitations
\cite{dau2,bang}. We did not show similar results that we have
obtained with this model, concerning the formation of a bubble out of thermal
excitation. However, this kind of process lacks any
possibility of control about the particular group of sites where it begins
to take place. Since it is certain that there is an enzymatic control
on the temporal and spatial beginning of the transcription, we have
adopted the point of view of mimicking in some way this initial action.
Another point to be noted is related to the energetics;
in the real process of transcription enzymes are present all the time;
this is in contrast with the strategy generally adopted, namely the study of
simple autonomous systems. However, one could argue that, if the autonomous
system shows dynamical processes that already enjoy a good degree of stability,
then the
enzymatic dynamical action (of course now we are not concerned with the
control activity), that should increase this stability and then
the lifetime of the process, requires a relatively small amount of energy.

At the beginning of the Introduction we pointed out that these models are
way too simple to represent faithfully objects as complex as, in general,
biological systems, and that their use is based, implicitly, on the
assumption, or better the hope, that their dynamical properties can
reproduce those of the real system, at least the more important. We think
that this point is strongly connected with the problem of the robustness,
previously mentioned with respect to changes within the framework of
the adopted model.
In fact, as long as one believes to have captured the essential properties
of the dynamics, one has also to be sure that an enrichment of the models,
necessary to get closer to more complete descriptions,
does not alter these properties. This is not a minor point: if the complexity
of the structure of a dynamical model increases, it probably becomes more
difficult to find a relatively ordered process as a travelling localized
excitation. We think that this is one of the problems that deserve the
efforts to be spent in future works.

\section*{ACKNOWLEDGMENTS}
It is a pleasure to acknowledge fruitful discussions with M. Barbi, S. Cocco
and A. Giansanti.

\appendix

\section{Equilibrium configurations in heterogeneous chains}
Let us suppose that we have found an equilibrium configuration for a
homogeneous chain with all $D_n$s equal to $D_{A-T}$. We now want to find
the configuration for a chain in which some of the $D_n$s are instead
equal to $D_{G-C}$. We can use the following procedure. We have to solve
Equations (\ref{eqconfa}) and (\ref{eqconfb}) for the given sequence of the
$D_n$s. We rewrite the equations in implicit form as:
\begin{equation}
\frac{\partial U}{\partial r_n}=0
\label{eqapa}
\end{equation}
\begin{equation}
\frac{\partial U}{\partial \theta_n}=0
\label{eqapb}
\end{equation}
Suppose to know the solution of (\ref{eqapa}) and (\ref{eqapb}) for a certain
sequence of the
$D_n$s. If we now have $D_n\rightarrow D_n+\delta D_n$, then we can find, at
the first order, the new solution by solving the linear system of equations:
\begin{equation}
\frac{\partial^2 U}{\partial r_n\partial D_n}\delta D_n+\sum_m
\frac{\partial^2 U}{\partial r_n\partial x_m}\delta x_m=0
\label{eqsha}
\end{equation}
\begin{equation}
\sum_m\frac{\partial^2 U}{\partial \theta_n\partial x_m}\delta x_m=0;
\,\,\,\,\,\,\,\,\,\,\,\,\,\,\, n=1,\ldots,N
\label{eqshb}
\end{equation}
where $x_m$ is the generic variable appearing in $U$, and in the sums in
$m$ the only terms that will appear are those belonging to the same
site or to the neighboring sites. In the linear system (\ref{eqsha}) and
(\ref{eqshb}) the
derivatives are to be computed in the old equilibrium configuration, and it
has to be solved with respect to the $\delta x_m$. Although this will give
the new configuration only at first order, it is nevertheless possible to
refine the solution up to the desired degree of accuracy with iterative steps.
If the values of $\frac{\partial U}{\partial r_n}$ and $\frac{\partial U}
{\partial \theta_n}$, after solving the system, are not yet equal to $0$
within the chosen tolerance, then one can solve a system
in which the terms with $\delta D_n$ in (\ref{eqsha})
are substituted by those values. With a suitable
choice for the variation $\delta D_n$ it is then possible, repeating a
sufficient number of times the procedure, to start from the solution of
(\ref{eqsha}) and (\ref{eqshb}) for $D_n=D_{A-T}$ for each $n$ and find the
solution in which any subset of the $D_n$s has become $D_{G-C}$.

\section{Equilibrium configurations with a bubble}
To solve in general Equations (\ref{eqconfa}) and (\ref{eqconfb}) we start
by posing:
\begin{equation}
\frac{L_{n+1,n}-L_0}{L_{n+1,n}}r_{n+1}r_n\sin\Delta\theta_n=s
\label{eqcons}
\end{equation}
In principle from (\ref{eqcons}) one can have $\Delta\theta_n=
f(r_{n+1},r_n,s)$. Substituting this function, for $\Delta\theta_n$ and
$\Delta\theta_{n-1}$, in (\ref{eqconfa}), one can obtain an equation
$g(r_{n+1},r_n,r_{n-1},s)=0$. Choosing the values of two contiguous sites
$r_m$ and $r_{m+1}$, the equilibrium configuration can be computed site by
site. But, without any hint on the choice of the initial values for $r_m$
and $r_{m+1}$, it is not possible to predict the structure along the chain
of the configuration that will be found. It would be the analogous of
computing a static solution of Eq. (\ref{fldeq}) with
``initial conditions'' on $\phi$ chosen at random: the solution $\phi(x)$
will be oscillatory or will (unphysically) diverge for $x\rightarrow +\infty$
or $x\rightarrow -\infty$; the localized solution such that $\phi(x)
\rightarrow \phi_{\pm}$ for $x\rightarrow \pm \infty$, where $\phi_+$ and
$\phi_-$ are two degenerate minima for $U$, requires exactly given ``initial
conditions''. We will show the way in which this problem can be solved and
therefore how we can find a solution of Equations (\ref{eqconfa}) and
(\ref{eqconfb})constituting a nontopological localized excitation.

As we said in subsection \ref{subeqcn}, posing $L_{n+1,n}=\overline{L}$
(constant in $n$) we find, from Eqs. (\ref{eqconfa}) and (\ref{eqconfb}), an
homogeneous equilibrium configuration with $r_n \equiv r$ and
$\Delta\theta_n=\Delta\theta$ for given $r$ and $\Delta\theta$; we consider
here the case $\overline{L}<L_0$, that gives $r>R_0$ and $\Delta\theta
<\Theta_0$. Substituting in (\ref{eqcons}) $r_{n+1}=r_n=r$ and
$\Delta\theta_n=\Delta\theta$, together with $L_{n+1,n}=\overline{L}$,
we find the corresponding value of $s$. With this value of $s$ we now want to
find a configuration such that $r_n\rightarrow r$ and $\Delta\theta_n
\rightarrow \Delta\theta$ for $n\rightarrow \pm \infty$, with an open region
in the middle. Then, for $n\rightarrow \pm \infty$ we write $r_n=r+\delta r_n$
and $\Delta\theta_n=\Delta\theta+\delta\theta_n$, we substitute in
(\ref{eqcons}) and we expand in power series of $\delta r_n$, $\delta r_{n+1}$
and $\delta\theta_n$, keeping only the first order terms. Therefore we have
a linear equation from which we obtain:
\begin{equation}
\delta\theta_n=\gamma\left(\delta r_n + \delta r_{n+1}\right)
\label{expan1}
\end{equation}
where the coefficient $\gamma$, that we do not write explicitly here, depends
on $r$, $\Delta\theta$ and the parameters of the model. At this point we
expand Eq. (\ref{eqconfa}) in power series of $\delta r_n$, $\delta r_{n+1}$,
$\delta r_{n-1}$, $\delta\theta_n$ and $\delta\theta_{n-1}$, we keep only
the first order terms and we substitute $\delta\theta_n$ and
$\delta\theta_{n-1}$ from (\ref{expan1}). Then we obtain $\delta r_{n+1}$
as a function of $\delta r_n$ and $\delta r_{n-1}$, in the form:
\begin{equation}
\delta r_{n+1}=\eta \delta r_n - \delta r_{n-1}
\label{expan2}
\end{equation}
Again, the coefficient $\eta$ depends on $r$, $\Delta\theta$ and the
parameters of the model. What is of interest here, for what will be said in a
moment, is that, when $r>R_0$ and $\Delta\theta<\Theta_0$, it is always
$\eta>2$. If we now pose:
\begin{equation}
\vec{\delta r}_n=\left(
\begin{array}{c}
\delta r_n \\ \delta r_{n-1} \\
\end{array} \right)
\end{equation}
then from (\ref{expan2}), we can derive the matrix equation:
\begin{equation}
\vec{\delta r}_{n+1}=\left(
\begin{array}{cc}
\eta & -1 \\
1 & 0 \\
\end{array}
\right) \vec{\delta r}_n
\label{matreq}
\end{equation}
The eigenvalues of the matrix in (\ref{matreq}) are given by:
\begin{equation}
\lambda_{\pm}=\frac{1}{2}\left[\eta \pm \sqrt{\eta^2-4}\right]
\label{eigv}
\end{equation}
The eigenvalues are both real and positive since $\eta>2$, and
$\lambda_-=\frac{1}{\lambda_+}$,
the matrix determinant being $1$. Then $\lambda_+>1$
and $\lambda_-<1$. We are therefore assured that, if for a given $n_0$ we
take $\vec{\delta r}_{n_0}$ as the eigenvector corresponding to
$\lambda_+$, then for $n<n_0$ $\delta r_n$ will tend exponentially to $0$.
Therefore, the strategy is to take such a $\vec{\delta r}_{n_0}$, obviously
of very small modulus to make the linear approximation in (\ref{eqcons}) and
(\ref{eqconfa})
valid, and then to compute $\Delta\theta_n$ and $r_n$ for $n>n_0$ site by
site from the
complete equations (\ref{eqconfa}) and (\ref{eqconfb}). One will reach a
maximum value of $r_n$
for some $n_1$, and for $n>n_1$ $r_n$ will decrease; only in the cases where
$r_{n_1-1}=r_{n_1+1}$ or $r_{n_1}=r_{n_1+1}$ a good localized solution, with
$r_n\rightarrow r$ for $n\rightarrow +\infty$, will be obtained. In the first
case we have an odd bubble, and in the second an even
bubble. To fall in one of these two cases, it is sufficient to
perform some tries adjusting the modulus of the initial vector
$\vec{\delta r}_{n_0}$.

In this way, we have found that in a somewhat uncoiled chain
($\Delta\theta_n<\Theta_0$) there is an equilibrium configuration with
$r_n\rightarrow r>R_0$ and $\Delta\theta_n\rightarrow \Delta\theta<\Theta_0$
for $n\rightarrow\pm\infty$, where in $r$ the hydrogen bond
represented by the Morse potential is only slightly stretched, and in the
middle $r_n$ is such that the hydrogen bond is completely broken (see Fig. 1).

As already explained, the qualitative picture is not changed in
heterogeneous chains.

\begin{figure}[p]
\includegraphics[129,130][566,630]{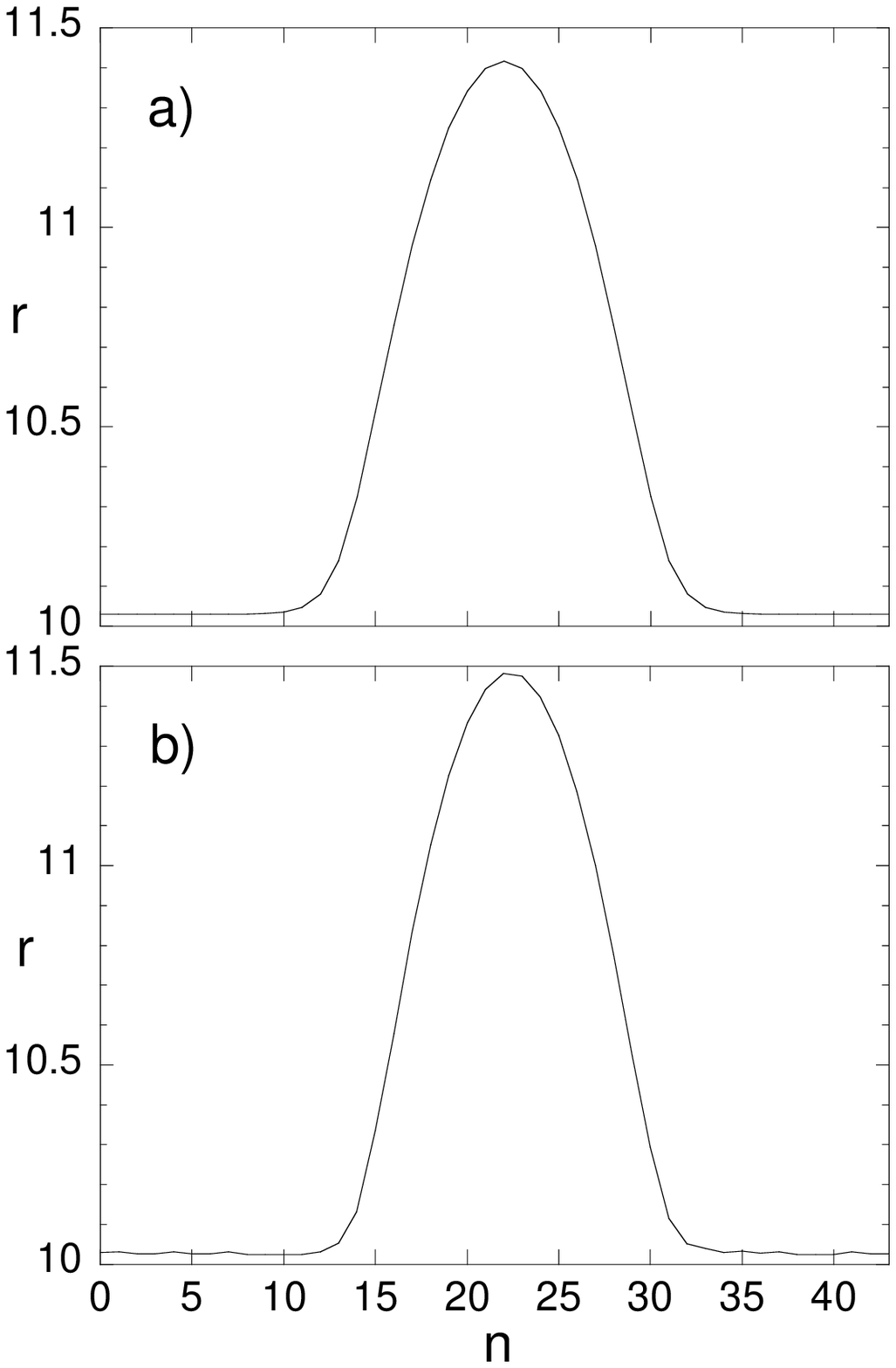}
\caption{Equilibrium configurations for a homogeneous chain with $D_n=D_{A-T}$
for each $n$ (a), and for a heterogeneous chain with a random choice of
$D_n$. We show only the central region, that with the bubble. In this and in
the figures from 3 to 7 the unit of length is \AA.}
\end{figure}
\begin{figure}[p]
\includegraphics[159,130][566,630]{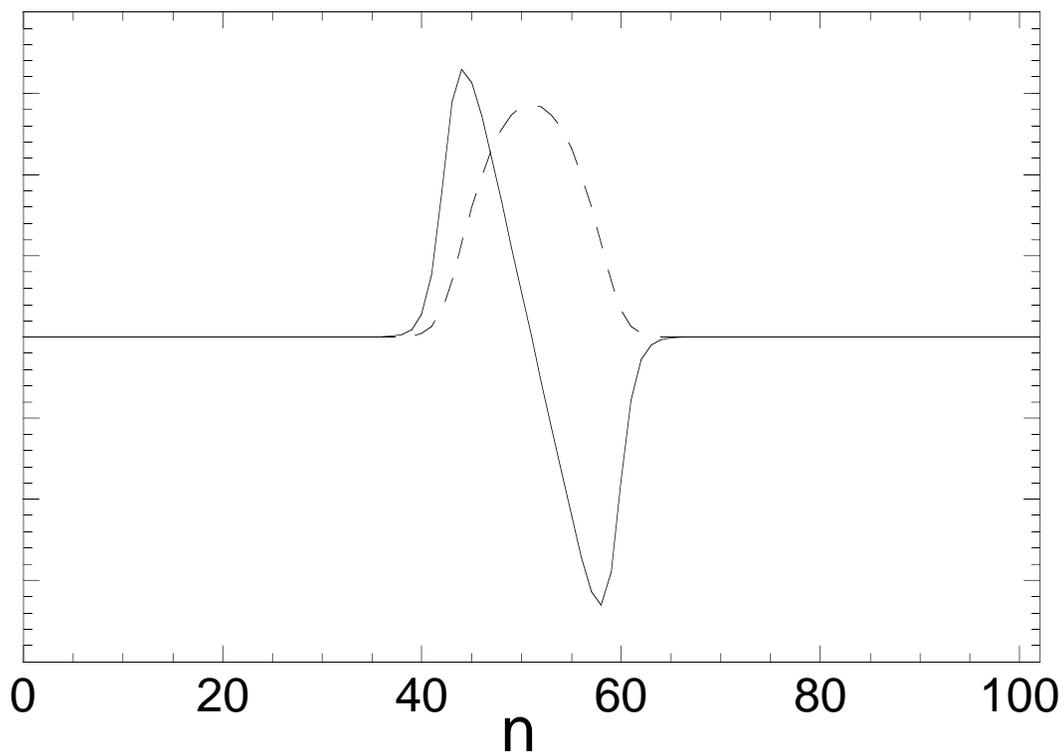}
\caption{Eigenvector of the smallest eigenvalue of the hessian matrix of
a homogeneous chain with $100$ base pairs. The full line corresponds to the
radial displacement $r$, while the dashed line to the transversal
displacement $R_0\theta$. In the vertical axis we have arbitrary units.}
\end{figure}
\begin{figure}[p]
\includegraphics[159,130][566,600]{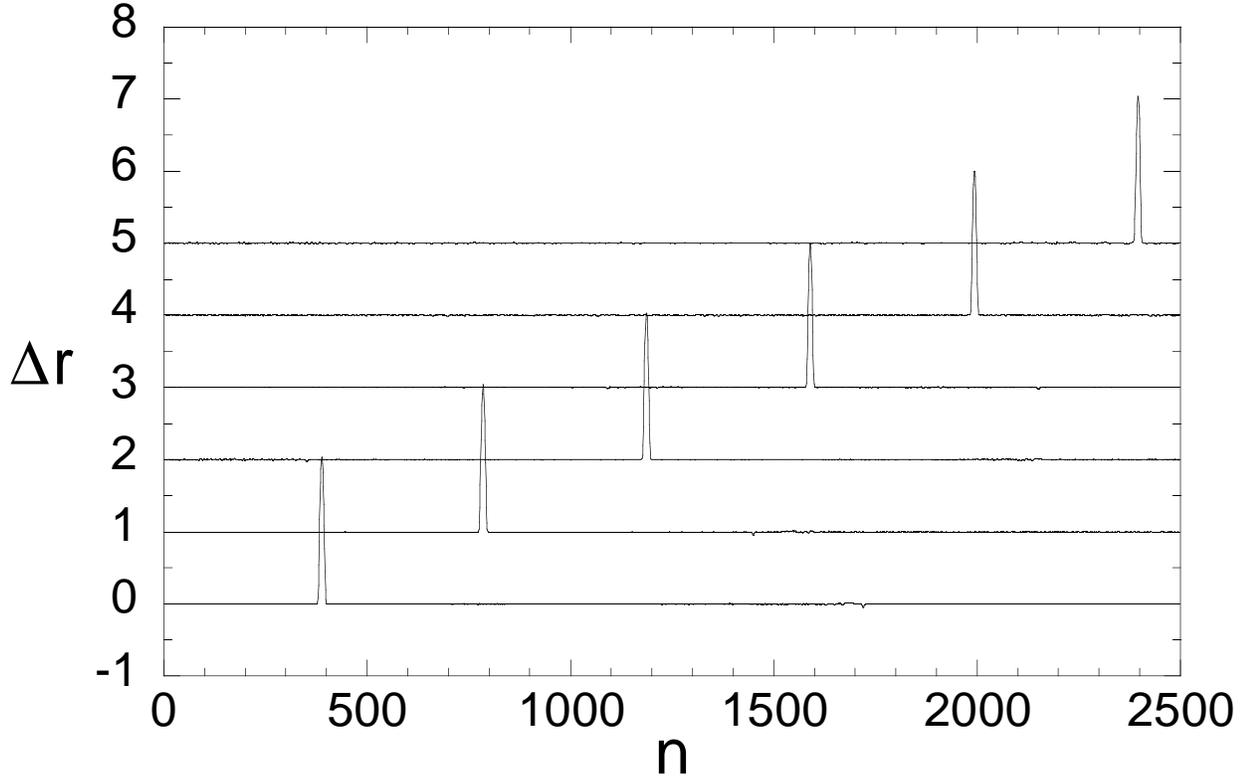}
\caption{Successive configurations in a homogeneous chain (all $D_n$s equal
to $D_{A-T}$) of $2500$ base pairs initially at rest in the fundamental
equilibrium configuration. In the vertical axis we have $\Delta r\equiv
r -R_0$. The bubble travels towards the right. It has been formed by an
unwinding given by an increase of $1.25$ radians in $\theta_0$. In order to
show all the configurations in a single graph, in the vertical position of
each one there is an offset of $1$ with respect to the previous.}
\end{figure}
\begin{figure}[p]
\includegraphics[159,130][566,600]{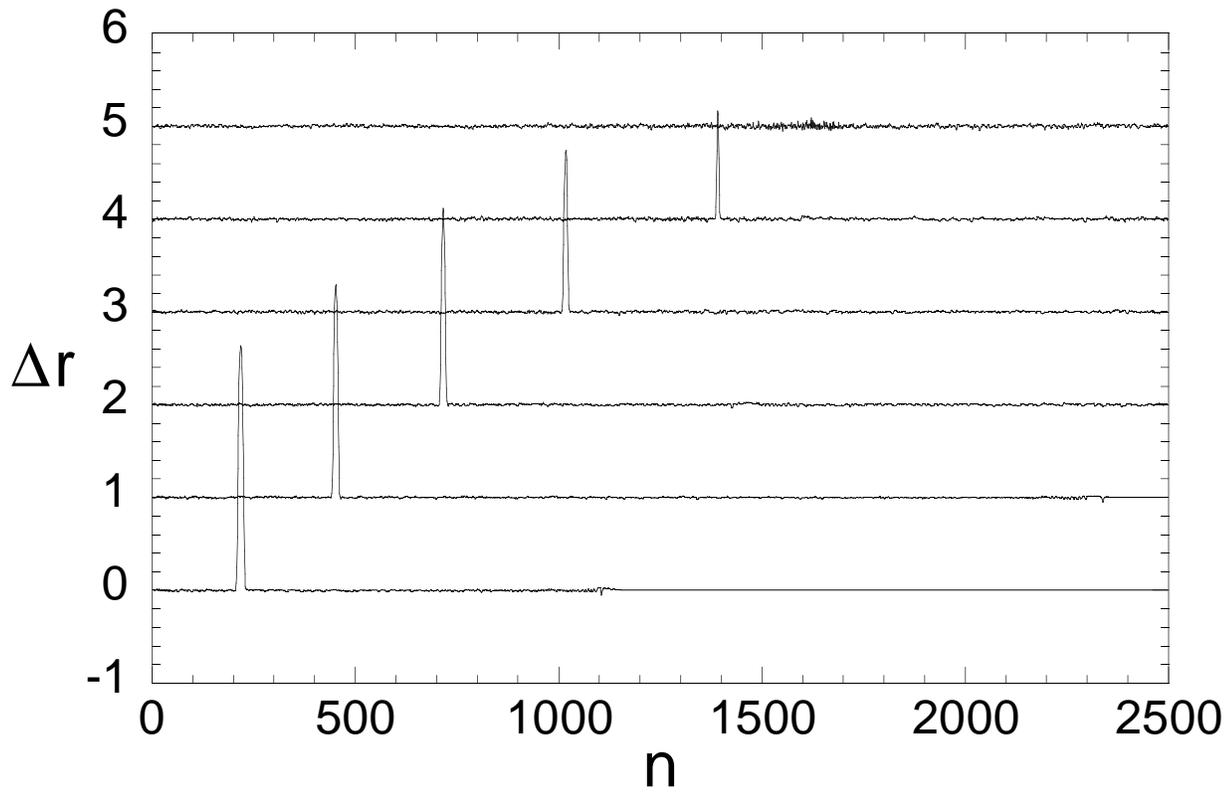}
\caption{Same as Fig. 3, but for a heterogeneous chain, with random choice of
the $D_n$s, and where $\theta_0$ has been initially increased by $2$ radians.}
\end{figure}
\begin{figure}[p]
\includegraphics[159,130][566,600]{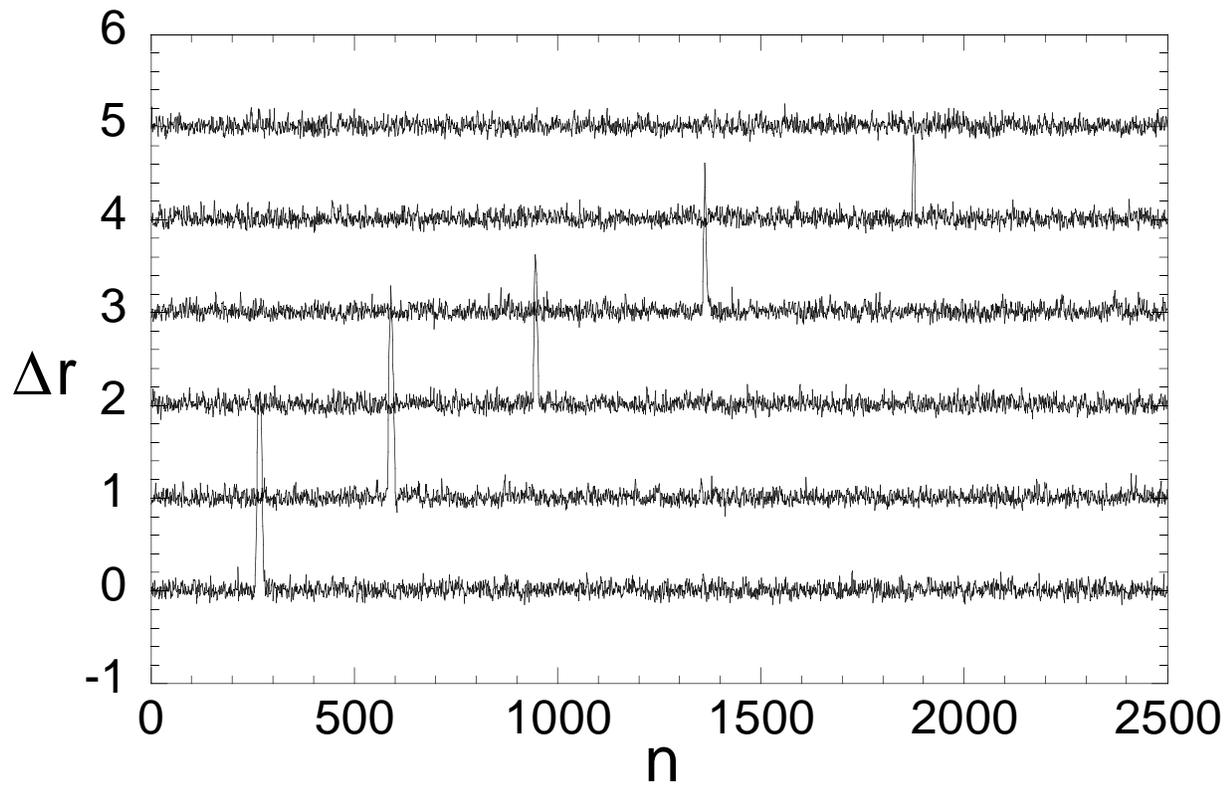}
\caption{Same as Fig. 3, but for a chain in thermal equilibrium at $300$ K,
and where $\theta_0$ has been initially increased by $2$ radians.}
\end{figure}
\begin{figure}[p]
\includegraphics[159,130][566,600]{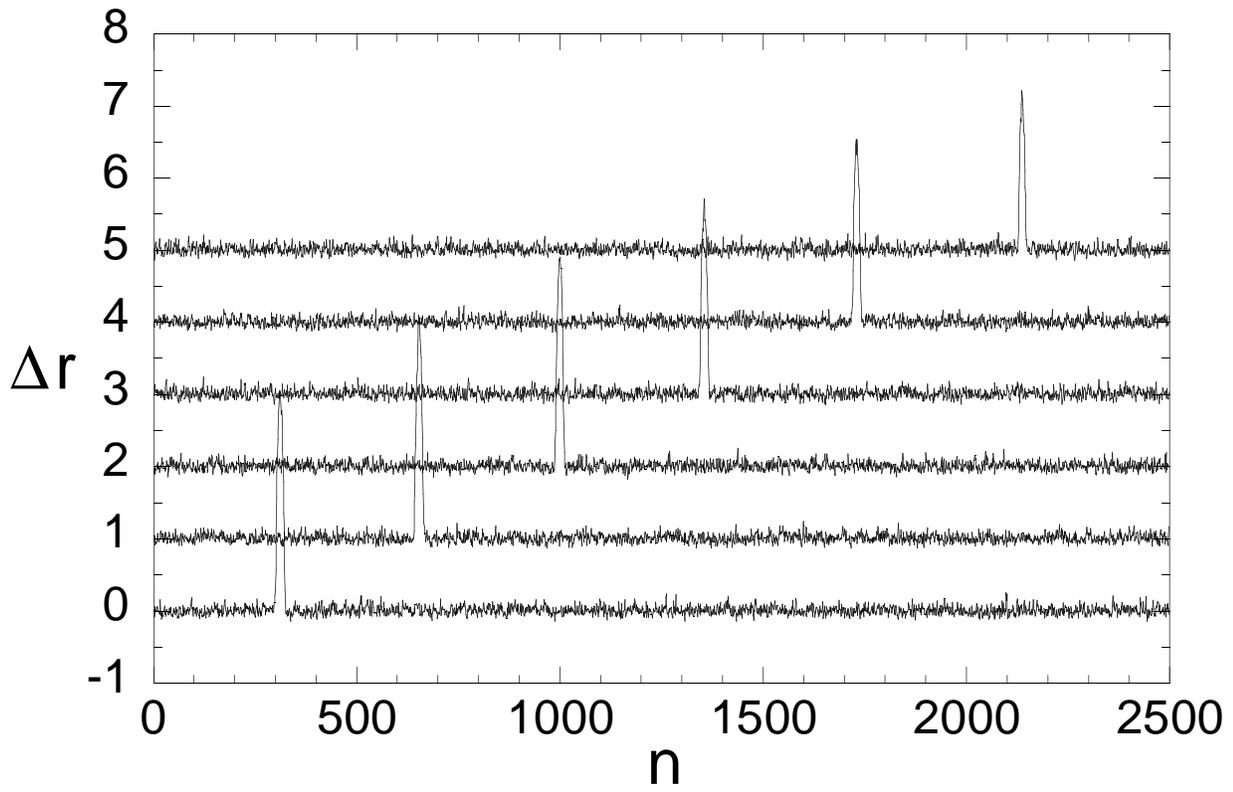}
\caption{Same as Fig. 5, but with $\theta_0$ initially increased by $2.8$
radians.}
\end{figure}
\begin{figure}[p]
\includegraphics[159,130][566,600]{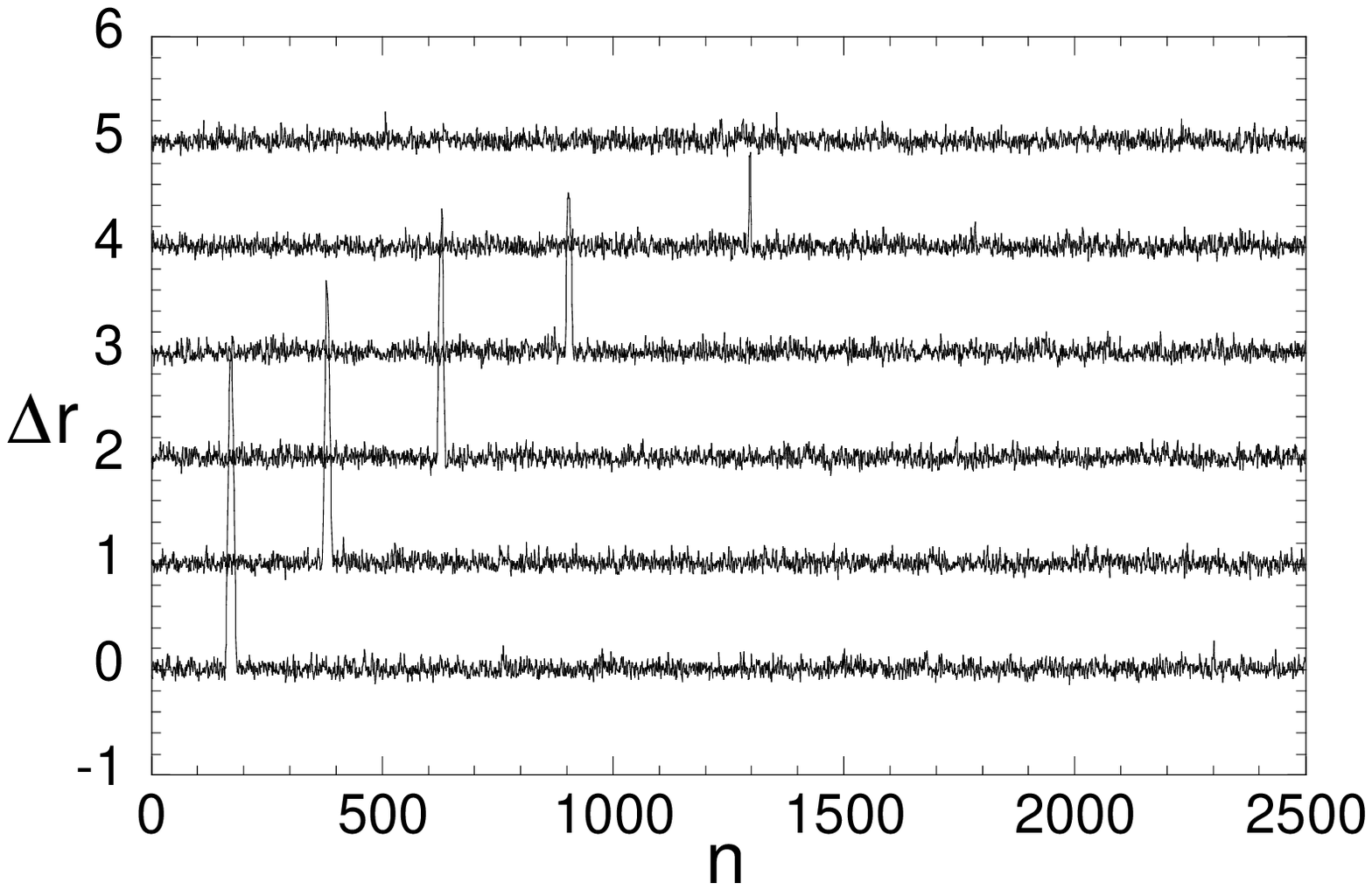}
\caption{Successive configurations in the same heterogeneous chain of Fig. 4,
but in thermal equilibrium at $300$ K and with $\theta_0$ initially increased
by $2.8$ radians.}
\end{figure}


\begin{thebibliography}{99}
\bibitem{not1}
Of course, with sufficiently long chains, the spectrum of phononic
perturbations will include arbitrarily small eigenfrequencies (associated
essentially to $\Delta\theta$ oscillations); however, this should not
modify the substance of our results: the number and the characteristics of
the eigenvectors associated to perturbations of the bubble should not change.
\end{thebibliography}
\end{document}